\documentclass[letterpaper, 10 pt, conference]{ieeeconf}  

\IEEEoverridecommandlockouts                              

\usepackage{graphicx} 
\usepackage{epsfig} 
\usepackage{amsmath} 

\title{\LARGE \bf
Comparison of Network Analysis Approaches on EEG Connectivity in Beta during Visual Short-Term Memory Binding Tasks
}

\author{Keith Smith$^{1,2}$, \textit{student member, IEEE},  Hamed Azami$^{1}$, \textit{student member, IEEE},\\ Javier Escudero$^{1}$, \textit{member, IEEE}, Mario A. Parra$^{2}$, John M. Starr$^{2}$ 
\thanks{*This work was partially supported by the Engineering and Physical Sciences Research Council. MAP was awarded an MRC Centenary Early Career Awards \#MRC-R42552. This work was conducted within the context of The University of Edinburgh Centre for Cognitive Ageing and Cognitive Epidemiology, part of the cross council Lifelong Health and Wellbeing Initiative (MR/K026992/1). MAP work is currently supported by Alzheimer's Society, Grant \#AS-R42303.}
\thanks{$^{1}$Keith Smith, Hamed Azami and Javier Escudero are with the Institute of Digital Communications, School of Engineering,
        University of Edinburgh,  King's Buildings, Edinburgh, UK, EH9 3JL.
        {\tt\small k.smith@ed.ac.uk,  hamed.azami@ed.ac.uk, javier.escudero@ed.ac.uk}}%
\thanks{$^{2}$Keith Smith, Mario A. Parra and John M. Starr are with Alzheimer Scotland Dementia Research Centre, University of Edinburgh,
		7 George Square, Edinburgh, EH8 9JZ.
        {\tt\small  mprodri1@staffmail.ed.ac.uk, jstarr@staffmail.ed.ac.uk}}%
}

\begin{document}

\maketitle
\thispagestyle{empty}
\pagestyle{empty}

\begin{abstract}

We analyse the electroencephalogram signals in the beta band of working memory representation recorded from young healthy volunteers performing several different Visual Short-Term Memory (VSTM) tasks which have proven useful in the assessment of clinical and preclinical Alzheimer's disease. We compare network analysis using Maximum Spanning Trees (MSTs) with network analysis obtained using 20\% and 25\% connection thresholds on the VSTM data. MSTs are a promising method of network analysis negating the more classical use of thresholds which are so far chosen arbitrarily. However, we find that the threshold analyses outperforms MSTs for detection of functional network differences. Particularly, MSTs fail to find any significant differences. Further, the thresholds detect significant differences between shape and shape-colour binding tasks when these are tested in the left side of the display screen, but no such differences are detected when these tasks are tested for in the right side of the display screen. This provides evidence that contralateral activity is a significant factor in sensitivity for detection of cognitive task differences. 

\end{abstract}

\section{Introduction}
Alzheimer's Disease (AD) is the most common form of dementia in the world and, due to the increasing age of the population, the number of people affected is likely to dramatically increase in the years to come. Visual Short-Term Memory  Binding (VSTMB) tasks are potentially useful in the detection of AD \cite{c1}\cite{c2}. AD patients perform significantly worse at shape-colour binding tasks than shape only tasks, whereas healthy old adults show no such diminished ability \cite{c10}. Further, no significant impairment is found in ability of VSTMB tasks due to non-AD dementias \cite{c1} and major depression \cite{c2}, suggesting specificity of impairment to AD. Still, there are many different and subtle factors which may affect cognitive tasks performance. Thus, we need to understand better the neurophysiological correlates of brain activity during these tasks in order to work towards a rigorous framework for assisting AD detection at preclinical stages. 

Analysis of Electroencephalogram (EEG) signals is particularly relevant for clinical detection of brain pathology in large at-risk populations, as required for AD. In this study we analyse EEG signal data of young healthy volunteers performing four distinct VSTM tasks, two of which involve memorising only shapes and the other two are VSTMB tasks, involving joint shape-colour binding memorisation. Here, we use network theory analysis in order to uncover differences affecting ability in these tasks.

Network theory is a widely applied analytical framework for studying interdependent phenomena \cite{c11} which is fast becoming a standard approach for complementing functional connectivity analysis in the brain \cite{c14}\cite{c15}. A network consists of a set of nodes with connections formed between them.  Network analysis is naturally suited to applications in EEG connectivity analysis where the electrodes form a bijective mapping with nodes in a network and the connections are defined by a similarity or dependency measure applied between pairs of EEG signals. The values acquired from these measures are in the form of an $n$x$n$ weighted adjacency matrix for a network with $n$ nodes. Applying the similarity measure between nodes $i$ and $j$ obtains the $i$th row and $j$th column entry of the adjacency matrix \cite{c11}. The networks obtained from similarity measures are complete, meaning connections exist between all possible pairs of nodes, and weighted, where measures give magnitudes between $0$ and $1$. Subjecting the resulting matrices to a threshold is desirable since it simplifies analysis and discards many uninformative low-weight connections.  However, there is currently no objective threshold for binarising the networks. This leads to study-by-study differences in choices leading to different and sometimes conflicting results \cite{c3}. 

A promising branch for unbiased network analysis is the Maximum Spanning Tree (MST) \cite{c4}\cite{c5}. A spanning tree is an acyclic, simple, connected, sub-network that connects to every node in the network and the maximum spanning tree is a spanning tree such that the sum of the weights of the connections in the original network included in the spanning tree are maximised \cite{c9}. This representation, when analysed in simulated networks, has been shown to be robust to underlying changes in the network and it overcomes the problem of threshold bias \cite{c5}. 

In this study we aim to use network analysis of EEG signals to look at differences occurring in brain activity when performing similar, but distinct visual short-term memory cognitive tasks. We want to understand more about these tasks in order to eventually test their application in detection of AD. In comparing the MSTs with standard threshold techniques, we wish to test suitability of MSTs for detecting changes in functional brain networks.  

\section{Materials}

\subsection{Subjects}

EEG signals were recorded for 23 healthy young volunteers participating in different VSTM tasks. Of the volunteers, five were left-handed and eight were women. Written consent was given by all subjects and the study was approved by the Psychology Research Ethics Committee, University of Edinburgh.

\subsection{Tasks} 

\begin{figure}[thpb]
	\centering
	\includegraphics[scale=0.23]{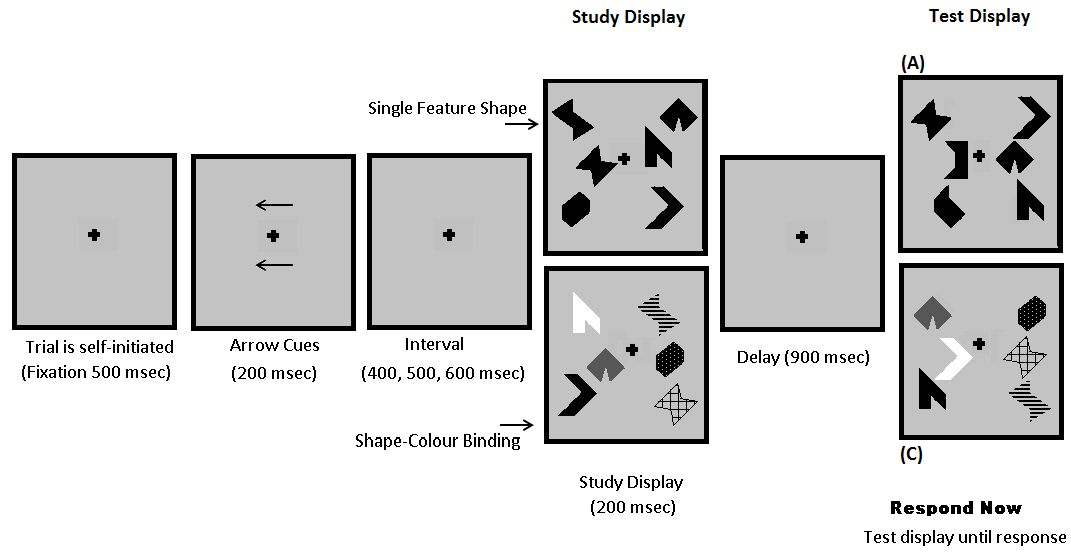}
	\caption{Structure of the VSTM tasks}
	\label{Taskfig}
\end{figure}

We consider a subset of four tasks of a larger study involving eight tasks. A schematic diagram of the tasks is shown in Fig. \ref{Taskfig} which also gives examples of the uncommon types of objects being probed. The positions of the objects were randomised separately for study and test displays to ensure that position was not a factor in memorisation. Each participant completed 8 practice trials followed by 170 test trials for each of the tasks. In half of the trials the objects of the study display were the same as the objects of the test display, while in the other half they were different. The test was then to decipher whether or not the objects in the study and test displays were the same which the volunteers indicated by pressing buttons with both hands. There were three objects in each hemisfield of the display screen. 
There are four distinct tasks distinguished by two binary conditions: i) single feature shape tested in left Hemisphere Response (HPR), ii) single feature shape tested in right HPR, iii) shape-colour binding tested in left HPR, iv) shape-colour binding tested in right HPR. Due to the contralateral behaviour of the brain, left and right hemisfield tests correspond to right hemisphere response (HPR) and left HPR, respectively.

\subsection{Recordings}

Only the trials with correct responses and no serious artefacts were kept since incorrect responses would not inform on memory binding activity. In a few cases, no useful data was available for a volunteer performing one of the tasks resulting in an unequal number of volunteers per task. 

We analysed the working memory representation, consisting of study display and maintenance periods, since this is found to be where differences in brain activity are most prevalent \cite{c18}. The EEG data was collected using NeuroScan version 4.3. This consisted of epochs with length of 1 second with -0.2 seconds of pre-stimulus recordings at 250 samples per second. A bandpass of 0.01-80 Hz was used in recording.
Forty EEG channels were recorded from common EEG sites, the majority of which were international 10/20 sites. Only thirty channels were kept for our purposes. The ten discarded channels consisted of four ocular channels, two linked mastoid reference channels and four which were discarded due to systematic noise (T5, T6, FT9 \& FT10). Fig. \ref{Mapfig} shows an abstract simplification of the EEG electrodes.

\section{Methods}

\subsection{Signal Processing}

Pre-processing, frequency analysis and connectivity analysis were performed using FieldTrip \cite{c6}. First, the 30 channels were re-referenced using an average reference which is more electrophysiologically silent \cite{c16}. Frequency analysis was then implemented from 0 seconds onwards using the multi-taper method with Slepian sequences and 2 Hz spectral smoothing. A one second zero-padding was applied to achieve 0.5 Hz resolution and the data was partitioned into five frequency bands. We focus here on $\beta$ (12.5 -32 Hz) due to its utility for AD detection \cite{c19} and sensory/memory integration \cite{c20}. After this, the debiased, Weighted Phase-Lag Index (dWPLI) \cite{c7}, an improved form of the Phase-Lag Index \cite{c8} for small sample sizes, was applied to obtain one connectivity matrix per trial. This similarity measure was chosen for its robustness to volume conduction effects as well as its ability to measure time-lagged inter-signal dependence which we assume is important for inter-regional communication in the brain \cite{c16}. 

\subsection{Network Theory}

\subsubsection{Maximum Spanning Trees}

From the resulting adjacency matrices, the MSTs were computed using an algorithm based on Kruskal's algorithm \cite{c12}. This adds the strongest weights in the network, as binary values, one by one to an empty $n$x$n$ matrix. At each step this MST matrix is checked for cycles. If no cycles are present, the next strongest connection is added and the algorithm continues. If a cycle is created from adding a connection, that connection is discarded before the algorithm continues. Cycles are checked using the property that, for a simple graph, $G$,
\[
G \text{ contains a cycle} \iff 0.5Trace(L)\geq Rank(L)+1,
\]
where $L$ is the Laplacian matrix of $G$. Once the MST matrix has $n-1$ connections with no cycles, i.e. is a spanning tree for the underlying network, the algorithm stops.
\begin{figure}[thpb]
	\centering
	\includegraphics[scale=0.2]{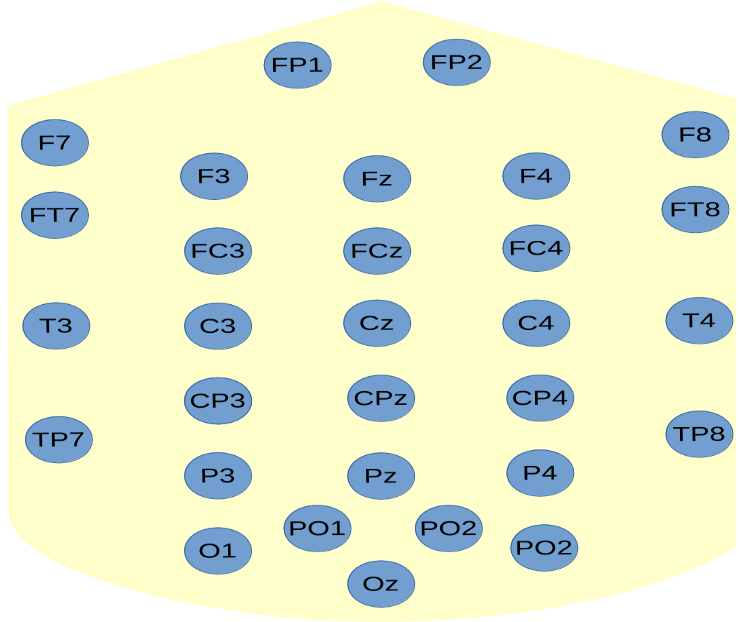}
	\caption{Simplified Map of channels.}
	\label{Mapfig}
\end{figure}

\subsubsection{Threshold Binarised Networks}

Threshold Binarised Networks (TBN) computed from thresholds keeping the 20\% and 25\% of strongest connections, rounded to the closest integer, were obtained for analysis and comparisons. Such thresholds are chosen suitably low to enhance comparability with MSTs.

\subsubsection{Whole Topology Network Measures}

We computed several common network measures for the whole topology of the networks. Correlations exist between the diameter and leaf fraction of the MST with the characteristic path length, $L$, and the local clustering coefficient, $C$, of TBNs, respectively \cite{c5}. We computed these values alongside the values of the maximum degree for both MSTs and TBNs. 

The Diameter of an MST is the longest shortest path between all pairs of nodes. This has an intuitive link to the characteristic path length, $L$, from standard network analysis, defined as the average shortest path length in the network, since a smaller diameter implies that there is a shorter route from any node to any other. The Leaf Fraction is the fraction of 1-degree nodes in the MST. The link to the local clustering coefficient, $C$, of standard networks, defined as average over all nodes of the probability that any two nodes each sharing a connection with a given node also share a connection, is less intuitive. Though it can partially be seen in that a high leaf fraction alludes to dense pockets of connectivity which implies higher clustering. The Maximum Degree of the network is the highest degree in the network where the degree of a node is the number of connections which that node shares in the network. 

\subsubsection{Node-specific Network Measures}

Noticing that, in terms of activity related to different tasks, the differences between networks may be quite subtle and so more obvious at a node-specific level, we also computed the eigenvector centrality. This measure gives high values for both high degree nodes and nodes connected to high degree nodes, thus measuring the importance of that node in the connectivity of the network \cite{c11}. This was computed for nodes O1, O2, P3 and P4 (see Fig. \ref{Mapfig}), since the parietal and occipital regions are instrumental regions for these tasks \cite{c18}. All of the measures were computed either using the Brain Connectivity Toolbox (BCT) \cite{c13} or else using straightforward calculations in MATLAB. 

	

\section{Results and Discussion}


\subsection{Whole Topology results}
   
A metric value was computed for the trial-average network of every volunteer during each of the tasks. These were then subjected to a two-way ANOVA test to look for differences between left and right HPR conditions, shape only and shape-colour binding conditions, and interaction between these conditions. A paired $t$-test was also run for shape only vs. shape-colour binding in left HPR and shape only vs. shape-colour binding in right HPR. We report here only on $p$-values which are significant at the 5\% level.

   
Notably, there were no significant differences found in the statistical analysis for the MST networks in the whole network measures.

\subsubsection{Two-way ANOVA} 
For the 20\% TBNs, significant interaction, $p=0.0100$, was found for $L$ between the left-right and shape-binding conditions. 
For the 25\% Threshold, significant difference was found in the Left-Right grouping for $C$ at $p=0.0184$, but none for interaction between Left-Right HPR and Shape-Bind conditions as found in the 20\% threshold case, particularly the $L$ $p$-value was above $0.1$.

\subsubsection{Paired \textit{t}-test}
For the 20\% TBNs, in shape vs. binding conditions for left and right HPR separately, it was found that every metric found significant differences in the right HPR conditions: $L$, $p=0.0243$; $C$, $p=0.0330$; Maximum Degree, $p=0.0405$. In contrast, there was no significant difference found in the left HPR conditions.
For the 25\% TBNs, again significant differences were found in shape vs. binding in the right HPR condition with none present in the left HPR condition. These were for $C$ at $p=0.0080$ and Maximum Degree at $p=0.0018$.
   
Since we assume that the functional connectivity networks during different task performances are different in reality, these results suggest that the MST is not a suitable method for detecting differences at the level of VSTM task performance in EEG signals. It may be that 30 electrodes is too few and that higher density EEG would prove beneficial for MST analysis. We found that some cases of noisy activity of a single node for a patient during one of the tasks proved to be a real problem for the MSTs. This resulted in cases with very high degree nodes, in some cases even hub like networks with one 29-degree node connected to twenty nine 1-degree nodes.  Noise removal attempts could be implemented to suppress this, however this would only introduce the arbitrary choices of threshold that the MST is put forward to avoid.
   
Furthermore, the MSTs may underrepresent strong regional activity in the network by the simple fact that no cycles and thus no clustering is allowed in the MST. Such activity may be vital in understanding the differences in activity between different VSTM task conditions. Likewise, MSTs may overrepresent weak activity in the network where connections which are much weaker than others may make it into the MST by virtue of the fact that the node which it connects to the network is underactive during the tasks.
   
TBNs prove more robust to noise simply due to the greater number of connections, and thus information, present. However, differences in results were apparent even when a fairly small increase of threshold was implemented, further exemplifying the threshold problem. All three metrics showed significant differences in right HPR in the 20\% case, whereas only $C$ and Maximum Degree showed significant differences in the 25\% case. Yet, the 25\% values were more significant than the 20\% values. Some of these may be explained by the fact that $L$ converges to $1$ as connections increase and $C$ and maximum degree converge to $0$ as connections decrease. 

We also adjusted our \textit{p}-values by implementing the false detection rate at $q=0.05$.  We note that the matter of dependency of different measures in networks for a given threshold is somewhat unclear, thus discretion in interpretation must be advised. Results indicated that only the values for $C$ and maximum degree in the 25\% TBNs are reliable results of a significant difference.
    
\subsection{Node-Specific Results}

The $p$-values from paired $t$-tests were evaluated for Eigenvector Centrality in nodes P3, P4, O1, and O2 for shape vs. shape-colour binding in the left and right HPR separately. 
\subsubsection{MSTs}
Similarly to the global network measures, no significant differences were found for the MST nodes. 
\subsubsection{20\% TBNs}
For the 20\% threshold, there was significant difference for P3 and O2 in the right HPR with $p = 0.0481$ and $p = 0.0417$, respectively. Interestingly, a significant difference was also found in the left HPR condition for P4 with $p = 0.0064$. 
\subsubsection{25\% TBNs}
For the 25\% Threshold, significant difference was found in right HPR for P3 with $p = 0.0443$. Signficant difference was found in left HPR for P4 with $p =0.0227$ and also in O2 with $p=0.0182$. The results found here agree for P3 and P4 but conflict with O2, again exemplifying threshold bias. 

Node specific metrics may be more helpful for detecting differences in left HPR. A more in depth analysis on a broader range of nodes would be helpful to understand these values and aid towards the framework of network theory as a tool for understanding interdependent functional activity in the brain during cognitive tasks.
Interestingly left HPR conditions show a difference in P4, in the right hemisphere of the brain, providing evidence that activity may be interdependent between hemispheres regardless of directed stimulation.

\section{Conclusions}

We found differences in network topologies of functional connectivity between performance of shape only and shape-colour binding tasks for right hemisphere response. This contrasted with a lack of evidence of differences for left hemisphere response, although at the level of node-specific measures some evidence was found for this. The right parietal region, node P4, was sensitive to differences in left HPR. Contralateral activity plays an important role in the VSTM tasks. Particularly, tasks for right hemisphere response may prove more sensitive to short-term memory as a potential test for AD. We also found that MSTs were unable to pick up on the differences clearly present in TBNs. Because of the small sample size of volunteers, these findings would benefit from more studies before conclusions can be drawn. The next step pertaining to MST sensitivity would be to keep information of weights for the connections in the MST to keep more information of the importance of connections. Further, an unbiased threshold is in development \cite{c21}.

\addtolength{\textheight}{-12cm}   






\end{document}